\documentclass{aastex63}
\usepackage{amsmath}
\usepackage{units}

\begin{document}
\title{The Next Generation Virgo Cluster Survey. XVII. A Search for Planetary Nebulae in Virgo Cluster Globular Clusters}

\correspondingauthor{Weijia Sun, Eric W. Peng}
\email{this.is.weijia@gmail.com, ewpeng@gmail.com}

\author[0000-0002-3279-0233]{Weijia Sun}
\affiliation{Department of Astronomy, Peking University, Yi He Yuan Lu 5, Hai Dian District, Beijing 100871, China}
\affiliation{Kavli Institute for Astronomy \& Astrophysics and Department of Astronomy, Peking University, Yi He Yuan Lu 5, Hai Dian District, Beijing 100871, China}

\author[0000-0002-2073-2781]{Eric W. Peng}
\affiliation{Department of Astronomy, Peking University, Yi He Yuan Lu 5, Hai Dian District, Beijing 100871, China}
\affiliation{Kavli Institute for Astronomy \& Astrophysics and Department of Astronomy, Peking University, Yi He Yuan Lu 5, Hai Dian District, Beijing 100871, China}

\author[0000-0001-6333-599X]{Youkyung Ko}
\affiliation{Department of Astronomy, Peking University, Yi He Yuan Lu 5, Hai Dian District, Beijing 100871, China}
\affiliation{Kavli Institute for Astronomy \& Astrophysics and Department of Astronomy, Peking University, Yi He Yuan Lu 5, Hai Dian District, Beijing 100871, China}

\author[0000-0003-1184-8114]{Patrick C{\^o}t{\'e}}
\affiliation{NRC Herzberg Astronomy \& Astrophysics Research Centre, 5071 West Saanich Road, Victoria, BC V9E 2E7, Canada}

\author[0000-0002-8224-1128]{Laura Ferrarese}
\affiliation{NRC Herzberg Astronomy and Astrophysics, 5071 West Saanich Road, Victoria, BC V9E 2E7, Canada}

\author[0000-0003-2713-6744]{Myung Gyoon Lee}
\affiliation{Astronomy Program, Department of Physics and Astronomy, Seoul National University, 1 Gwanak-ro, Gwanak-gu, Seoul 08826, Korea}

\author[0000-0002-4718-3428]{Chengze Liu}
\affiliation{Department of Astronomy, Shanghai Key Laboratory for Particle Physics and Cosmology, Shanghai Jiao Tong University, China}

\author[0000-0001-5569-6584]{Alessia Longobardi}
\affiliation{Aix Marseille Universite{\'e}, CNRS, CNES, Laboratoire d'Astrophysique de Marseille UMR 7326, F-13388, Marseille, France}

\author[0000-0002-7924-3253]{Igor V. Chilingarian}
\affiliation{Harvard-Smithsonian Center for Astrophysics, 60 Garden St. Cambridge MA 02138, USA}
\affiliation{Sternberg Astronomical Institute, M.V.Lomonosov Moscow State University, Universitetsky prospect 13, Moscow, 119234, Russia}

\author[0000-0002-1685-4284]{Chelsea Spengler}
\affiliation{Institute of Astrophysics, Pontificia Universidad Católica de Chile, Av. Vicuña Mackenna 4860, 7820436 Macul, Santiago, Chile}
\affiliation{Kavli Institute for Astronomy \& Astrophysics and Department of Astronomy, Peking University, Yi He Yuan Lu 5, Hai Dian District, Beijing 100871, China}
\affiliation{Department of Astronomy, Peking University, Yi He Yuan Lu 5, Hai Dian District, Beijing 100871, China}

\author[0000-0001-6047-8469]{Ann I. Zabludoff}
\affiliation{Steward Observatory, University of Arizona, 933 North Cherry Avenue, Tucson, AZ 85721, USA}

\author[0000-0003-1632-2541]{Hong-Xin Zhang}
\affiliation{CAS Key Laboratory for Research in Galaxies and Cosmology, Department of Astronomy, University of Science and Technology of China, Hefei, Anhui 230026, China}
\affiliation{School of Astronomy and Space Science, University of Science and Technology of China, Hefei 230026, China}

\author{Jean-Charles Cuillandre}
\affiliation{CEA/IRFU/SAp, Laboratoire AIM Paris-Saclay, CNRS/INSU, Universit{\'e} Paris Diderot, Observatoire de Paris, PSL Research University, F-91191 Gif-sur-Yvette Cedex, France}

\author{Stephen D. J. Gwyn}
\affiliation{NRC Herzberg Astronomy and Astrophysics, 5071 West Saanich Road, Victoria, BC V9E 2E7, Canada}

\begin{abstract}
The occurrence of planetary nebulae (PNe) in globular clusters (GCs) provides an excellent chance to study low-mass stellar evolution in a special (low-metallicity, high stellar density) environment. We report a systematic spectroscopic survey for the [O{\sc iii}] \unit[5007]{\AA} emission line of PNe in 1469 Virgo GCs and 121 Virgo ultra-compact dwarfs (UCDs), mainly hosted in the giant elliptical galaxies M87, M49, M86, and M84. We detected zero PNe in our UCD sample and discovered one PN ($M_{5007} = \unit[-4.1]{mag}$) associated with an M87 GC. We used the [O{\sc iii}] detection limit for each GC to estimate the luminosity-specific frequency of PNe, $\alpha$, and measured $\alpha$ in the Virgo cluster GCs to be $\alpha \sim \unit[3.9_{-0.7}^{+5.2}\times 10^{-8}]{PN/\mathit{L}_\odot}$. $\alpha$ in Virgo GCs is among the lowest values reported in any environment, due in part to the large sample size, and is 5--6 times lower than that for the Galactic GCs. We suggest that $\alpha$ decreases towards brighter and more massive clusters, sharing a similar trend as the binary fraction, and the discrepancy between the Virgo and Galactic GCs can be explained by the observational bias in extragalactic surveys toward brighter GCs. This low but non-zero efficiency in forming PNe may highlight the important role played by binary interactions in forming PNe in GCs. We argue that a future survey of less massive Virgo GCs will be able to determine whether PN production in Virgo GCs is governed by internal process (mass, density, binary fraction), or is largely regulated by external environment.
\end{abstract}
\keywords{galaxies: individual (M87, M49, M86, M84) -- globular clusters: general -- planetary nebulae: general -- stars: evolution}

\section{Introduction \label{sec:intro}}

In spite of decades of research, the stellar population that forms PNe is still not clear. Traditionally, PNe are thought to be the circumstellar gas ejected from extreme giant stars at the end of the asymptotic giant branch (AGB) phase, accompanied with an intense mass loss \citep{1966PASP...78..232A}. In this view, nearly all stars with a mass between \unit[1.2]{$M_\odot$} and \unit[8]{$M_\odot$} eventually become PNe. However, there remain challenges in explaining the axisymmetric and point-symmetric morphology of PNe, especially young ones \citep{1998AJ....116.1357S}. Stellar rotation and magnetic fields \citep{1999ApJ...517..767G, 2005ApJ...618..919G} have been proposed to explain this phenomenon, but both suffer from fundamental defects \citep{2007MNRAS.376..599N, 2014ApJ...783...74G}. Another promising theory involving central star (CS) binarity might be the key to understand PNe formation \citep[e.g.,][]{1997ApJS..112..487S, 2005ApJ...629..499C, 2006ApJ...645L..57S, 2009PASP..121..316D, 2017NatAs...1E.117J}. The mass-transfer between a close binary can lead to a bipolar outflow \citep{2006MNRAS.370.2004N} and prolong strong magnetic fields by providing extra angular momentum \citep{2014MNRAS.439.2014T}. Since the first PN with a confirmed binary CS was discovered \citep{1976PASP...88..192B}, more than 60 binary CSs have been found\footnote{\url{http://www.drdjones.net/bCSPN/}}, laying a solid foundation for a binary interaction model.

GCs provide a unique environment to test this scenario. Stars of old populations (like those in GCs) with turnoff mass $\sim \unit[0.8]{\mathit{M}_\odot}$ will have $\sim \unit[0.5-0.53]{\mathit{M}_\odot}$ \citep{2000AJ....120.2044A, 2009ApJ...705..408K} remaining at the end of AGB phase. The evolutionary timescale is longer than \unit[$10^5$]{yr} \citep{1989A&A...221...27C} for such a low-mass core, thus any remnant ejected at the end of the AGB phase would have dispersed before the CS gets hot enough to ionize the expelled gas, making it impossible to form any PNe \citep{1983ApJ...272..708S}. \citet{2006MNRAS.368..877B} also argued that, below a critical mass of $\sim \unit[0.52]{\mathit{M}_\odot}$, no PN events are expected to occur under the assumption of single-star evolution. Therefore, any detection of PNe in GCs suggests an alternative evolutionary channel. The binary interaction model may provide a plausible explanation because a common-envelope interaction or even a stellar merger, which is favored by GCs' high-density environment, will accelerate the evolutionary process and likely form a PN with low-mass CSs.

Since the first PN belonging to a GC was found in 1928 in M15 \citep{1928PASP...40..342P}, just three more PNe in GCs have been detected in the Milky Way (MW) \citep{1989ApJ...338..862G, 1997AJ....114.2611J}. Meanwhile, there have been some serendipitous discoveries for PNe in extragalactic GCs \citep{2002ApJ...575L..59M, 2006AA...448..155B, 2008AA...477L..17L, 2008AJ....136..234C}. \citet{2012ApJ...752...90P} conducted a systematic survey of the giant elliptical galaxy NGC 4472 in the Virgo cluster using FLAMES/GIRAFFE at the Very Large Telescope and found zero PNe in 174 GCs. Another ground-based spectroscopic survey in M31 \citep{2013ApJ...769...10J} discovered 3 candidate PNe in 274 GCs. However, \citet{2015AJ....149..132B} ruled out one object found by \citet{2013ApJ...769...10J} through \textit{Hubble Space Telescope} snapshot imaging. The luminous [O{\sc iii}] emission makes it possible to detect PNe at a distance beyond the Milky Way via a narrow band filter centered on the redshifted wavelength of the [O{\sc iii}] line \citep{1997MNRAS.284L..11T, 1990ApJ...356..332J} or using medium-resolution spectrograph. Low-mass X-ray binaries are another potential source of [O{\sc iii}] emission in GCs. X-ray and [O{\sc iii}] emission has previously been observed from a GC in NGC4472 \citep{2007ApJ...669L..69Z} and in NGC 1399 \citep{2010ApJ...712L...1I}. These low-mass X-ray binaries can be distinguished from PNe through their broad [O{\sc iii}] emission line profiles.

The number of PNe in a stellar population per unit bolometric solar luminosity, $\alpha$, has been used to place a constraint on the proportion of stars that contribute to planetary nebula luminosity function (PNLF) cut-offs. Galaxies with integrated $B-V$ smaller than \unit[0.8]{mag} are characterized by $\alpha \sim \unit[3\times 10^{-7}]{PN/\mathit{L}_\odot}$, with the spread in $\alpha$ increasing significantly as the color becomes redder with respect to the constant value observed in bluer objects \citep{1990RPPh...53.1559P, 2006MNRAS.368..877B}. Measurements for $\alpha$ in GCs are limited. In Galactic GCs, $\alpha \sim \unit[2.0\times 10^{-7}]{PN/\mathit{L}_\odot}$ and, in NGC 4472, \citet{2012ApJ...759..126P} gave an upper limit of $\alpha \leq \unit[8\times 10^{-8}]{PN/\mathit{L}_\odot}$ based on a non-detection.

Given that the mean GC luminosity is $\sim 10^5L_\odot$, hundreds of GCs need to be surveyed to place interesting constraints on $\alpha$ in these environments. Here we report a systematic survey for PNe within GCs in the Virgo cluster with a sample size that is more than 5 times larger than the previous largest survey \citep{2013ApJ...769...10J}. In Section \ref{sec:data}, we describe our data and reduction procedures. In Section \ref{sec:find}, we describe our PN detection and confirmation procedures. Then in Section \ref{sec:discussion}, we discuss the choice of PNLF and we use the detection limits for each GC to quantify their contribution to the estimation of $\alpha$. We also discuss implications for PN formation channels. Finally, we summarize our study in Section \ref{sec:conclusion}.

\section{Data Reductions \label{sec:data}}

This work is based on the spectroscopic data of GC candidates around the giant elliptical galaxies M87, M49, M86 and M84 in the Virgo cluster, which were collected as part of the Next Generation Virgo Cluster Survey (NGVS). It offers deep imaging data of the Virgo cluster within its virial radius ($\sim\unit[104]{deg^2}$) using the MegaCam instrument on the Canada-France-Hawaii Telescope \citep{2012ApJS..200....4F}. The selection of GCs around M87 and the reduction of their spectra were described in \cite{2015ApJ...802...30Z, 2017ApJ...835..212K}. The primary selection for our GC targets is based on color. We also considered the concentration index as a secondary selection criterion. Since most of the GCs are unresolved at the distance of the Virgo cluster, it only biases the GC selection marginally based on their densities or sizes. The authors used the Hectospec multifiber spectrograph \citep{2005PASP..117.1411F} on the \unit[6.5]{m} MMT telescope to carry out an extensive spectroscopic survey of Virgo GCs and UCDs (\unit[3650 -- 9200]{\AA}, $R \sim 1000$). The data were reduced through version 2.0 of HSRED reduction pipeline\footnote{\url{http://www.mmto.org/node/536}}, which includes the basic corrections (bias, dark and flat-fielding), aperture extraction, wavelength calibration. We did the flux calibration following the steps described in \citet{2008PASP..120.1222F}, which include the corrections for atmospheric extinction, Hectospec relative throughput, and absolute flux normalization with MegaCam $g^\prime$ photometry. The radial velocities were determined by a cross-correlation with SDSS stellar templates using the IRAF package RVSAO\footnote{\url{http://tdc-www.harvard.edu/iraf/rvsao}} \citep{1998PASP..110..934K}. As for the GCs around M49, they were collected and reduced following similar methods as for the GCs in M87 (Ko, Y. et al., in prep.). The median signal-to-noise ratio (SNR) per pixel is around 10, with $\mathrm{SNR} > 5$ for more than 80\% of our sample. The radial velocity uncertainties were derived from the correlation error, and their median velocity uncertainty is around $\unit[31]{km\,s^{-1}}$. In the following analysis, the spectra were shifted to the rest frame on the basis of the derived radial velocities correspondingly.

 To avoid the contamination of foreground stars or background galaxies, we distinguished the GCs from contaminants based on their radial velocities and photometric information including concentration indices defined as the difference between \unit[4]{pix} and \unit[8]{pix} aperture-corrected magnitude, half-light radii, and colors. First, all the spectroscopic targets with $v_\mathrm{r} >\unit[3000]{km\,s^{-1}}$ were considered background sources \citep[see references in][]{2017ApJ...835..212K}. Second, we distinguished ultra-compact dwarfs (UCDs) among our sample. We measure the half-light radii for the bright sources with $i < \unit[21.5]{mag}$. The sources with half-light radii of $r_\mathrm{h}>\unit[11]{pc}$ are considered as the UCDs. Third, for the remaining sources, we separate them into two groups through radial velocities. For the sources with radial velocities higher than $\unit[500]{km\,s^{-1}}$, we consider them as the GCs, because the Galactic stars rarely have those high radial velocities. For the sources with lower radial velocities, we adopt the extreme deconvolution algorithm \citep{2011ApJ...729..141B} on the multi-dimensional color space and the concentration index \citep[see Section 2.3.1 in][]{2018ApJ...864...36L}. Based on this process, we remove foreground Galactic stars from our sample. 

\begin{figure*}[ht!]
\plotone{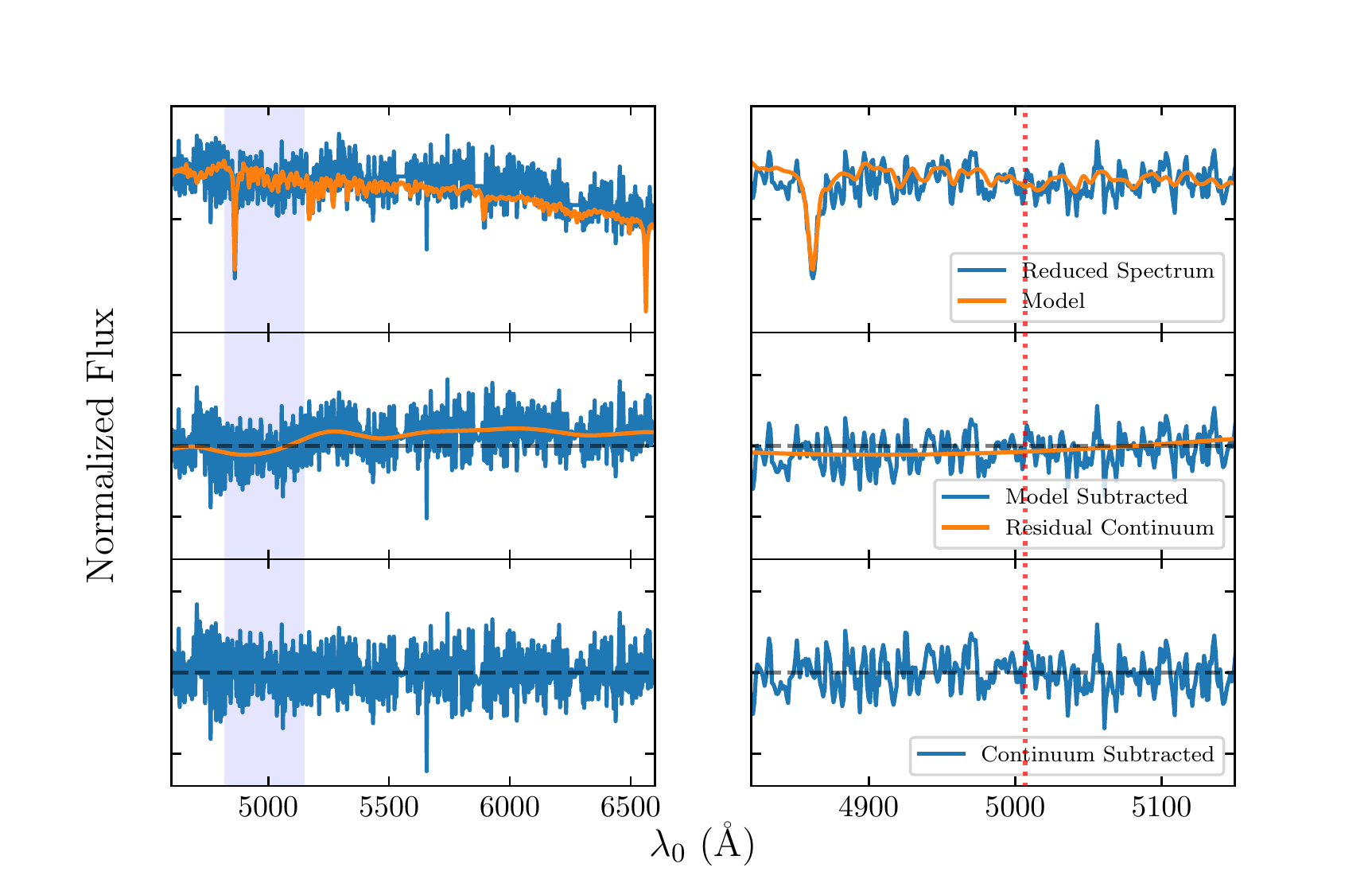}
\caption{One representative spectrum (in the rest frame) of an M87 globular cluster, showing the reduced spectra (blue in the top panel), along with the best fit model (orange in the top panel), the spectra after model subtraction (blue in the middle panel) and those after residual continuum subtraction (blue in the bottom panel). The right panel shows a zoomed-in view of the shadowed area in the left panel, covering two different wavelength ranges for the same object. The vertical dashed line indicates the location of \unit[5007]{\AA}. \label{fig:reduction}}
\end{figure*}

Instead of using five templates to detect the emission line features \citep{2013ApJ...769...10J}, we used simple stellar population (SSP) models of \citet[][hereafter BC03]{2003MNRAS.344.1000B} adopting the \citet{2003PASP..115..763C} IMF with metallicity from $Z = 0.0001$ to $0.05$ and age from \unit[1 -- 10]{Gyr}. Ages and masses for these clusters were determined by comparison with models \citep[e.g.,][]{1994ApJS...95..107W}. The BC03 models do not include the emission lines of a possible PN, thus a model-subtracted spectrum should preserve these emission lines. Some redundant continuum components remained after the subtraction of the model spectrum (see the middle panels of Fig.~\ref{fig:reduction}). Since they had little or no influence on the emission line features we were interested in, they were subtracted before we performed a $\chi^2$ test to derive the best-fit model. The calibrated spectra are present in the bottom panel of Fig.~\ref{fig:reduction} with a vertical dashed line indicating the location of \unit[5007]{\AA}. Through visual inspection, we verified this procedure works well for most of our clusters.

\section{Finding planetary nebulae in globular clusters \label{sec:find}}
To verify the detection of PNe in GCs, we used three criteria. The emission line of [O{\sc iii}] \unit[5007]{\AA} is the most notable feature of a PN, which makes it the most important criterion. In the second place, emission line profiles and flux ratios between the emission lines ([O{\sc iii}] \unit[4959]{\AA}, \unit[5007]{\AA} and H$\alpha$/H$\beta$) should be consistent with those expected from a PN. Last but not least, the radial velocity of a candidate, from its emission lines, must be consistent with its parent GC. 

We searched for lines whose flux deviation from zero is larger than \unit[3]{$\sigma$} where $\sigma$ is estimated from a part of the calibrated spectrum where no strong absorption or emission lines exist. If an emission line feature is spotted within \unit[3]\AA around \unit[5007]{\AA}, this might be a candidate PN. Note that the value of the typical error in redshift determination is around $\unit[31]{km\,s^{-1}}$ (corresponding to an error of \unit[0.5]{\AA} in wavelength), therefore it will not effect the detection for the [OIII] 5007 emission from PNe. We also looked for the emission lines around \unit[4959]{\AA}, which is another noteworthy feature of PNe. The flux ratios among the emission lines have been proven to be an effective tool in identifying PNe. \citet{2003AJ....125..514A} first used $I(\lambda 4959)/I(\lambda 5007)$ to confirm extragalactic PNe in the Virgo cluster. The flux ratio between \unit[4959]{\AA} and \unit[5007]{\AA} should be close to one third ($I(\lambda 4959)/I(\lambda 5007)\approx 1/ 3$) and the [O{\sc iii}] \unit[5007]{\AA} to H$\beta$ ratio should larger than 2 ($I(\lambda 5007)/I(\mathrm{H}\beta) \geqslant 2$) in the metal-poor environments of GCs \citep{2013ApJ...769...10J}. For most of our clusters, however, reliable detection of Balmer lines and the determination of flux ratio may suffer from low signal-to-noise ratio. The width of $\lambda 5007$ should be identical to the spectral resolution, excluding the chance of being a supernova remnant or X-ray binary, which has much broader emission lines than any PN \citep{1980ApJ...235..939W}.

\begin{figure*}[ht!]
\plottwo{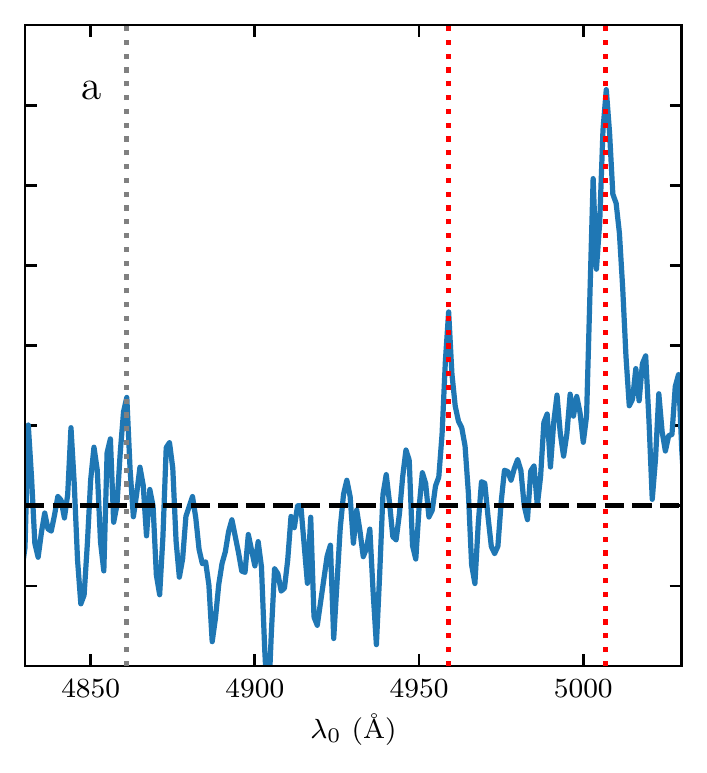}{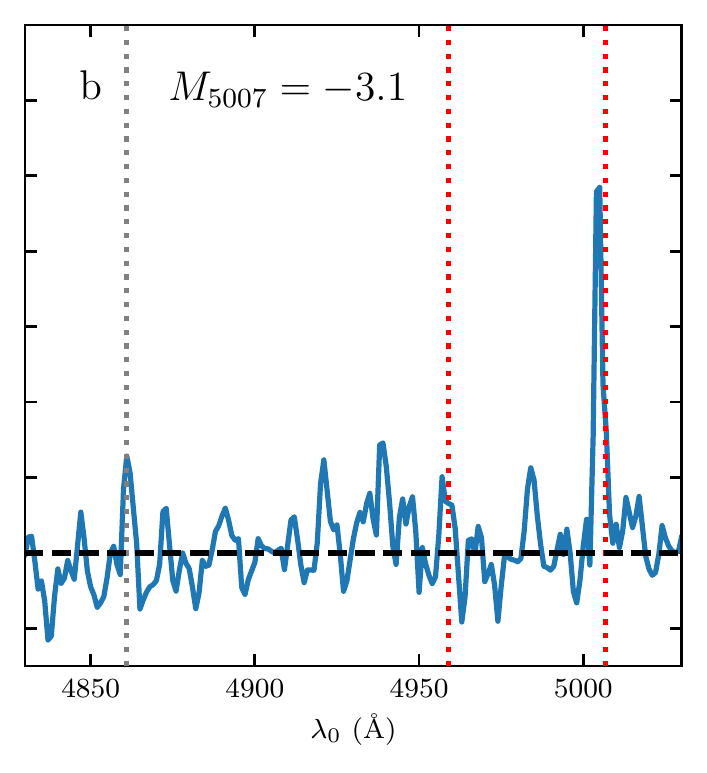}
\plottwo{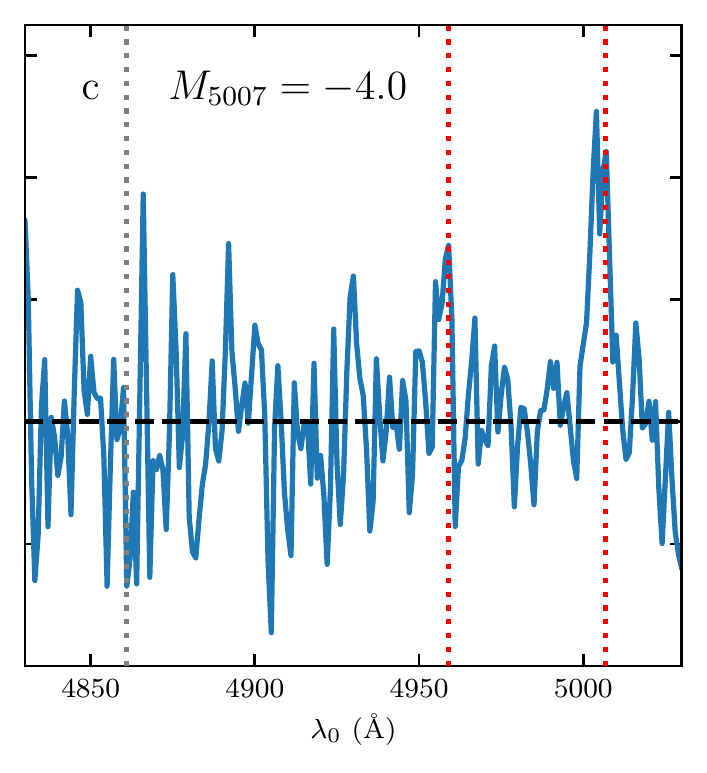}{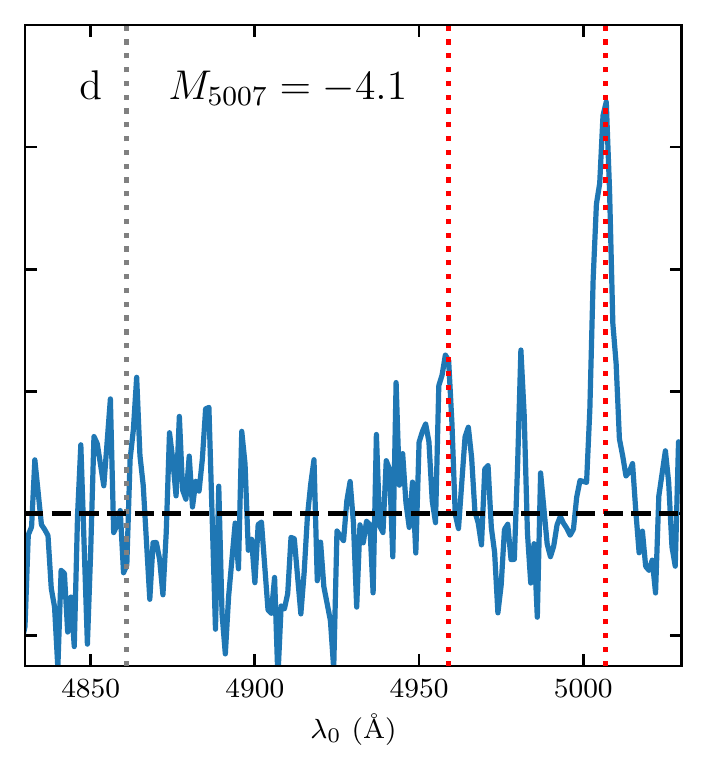}
\caption{The spectra of four GCs with candidate PNe with emission lines around [O{\sc iii}] \unit[4959]{\AA}, \unit[5007]{\AA} with its identifier and the absolute magnitude of \unit[5007]{\AA} emission (if available) in the middle of each figure. $M_\mathrm{5007}$ of GC-c is calculated by summing up the flux in both peaks. The red vertical dashed lines represent [O{\sc iii}] \unit[4959]{\AA}, \unit[5007]{\AA} and the grey vertical dashed line represents H$\beta$ in the rest frame. The black horizontal dashed lines represent zero flux. \label{fig:spec}}
\end{figure*}

To estimate the absolute magnitude of [O{\sc iii}] $\lambda 5007$, we first inferred the continuum brightness of each cluster from the MegaCam $g^\prime$ band magnitude. The monochromatic flux around \unit[5007]{\AA} was then estimated by comparing the strength of the emission relative to the continuum, i.e., the equivalent width. This corrects for flux loss from the fiber aperture. After that, the flux was transformed into apparent [O{\sc iii}] $\lambda 5007$ magnitude according to the relationship adopted in \citet{1989ApJ...339...53C}:
\begin{equation}
m_{5007} = -2.5 \log F_{5007} - 13.74
\end{equation}
In this way, we built a connection between $m_{5007}$ and $m_{g^\prime}$ using the flux ratio between the emission line and the continuum, which obviates the problem of flux calibration for fiber spectrophotometry. GCs are classified into two groups, adopting the criteria used in \citet{2018ApJ...864...36L}. Briefly, If the GCs are located within 10 effective radii, $R_\mathrm{e}$, of the galaxy and have the relative radial velocities to the galaxy within $3\sigma_g$ where $\sigma_g$ is the galaxy velocity dispersion, they are considered members of that galaxy. The effective radius and velocity dispersion information of Virgo galaxies were adopted from the NGVS galaxy catalog (Ferrarese et al., in prep.) and the values compiled by the GOLDMine project, respectively. If there is no information about the radius and velocity dispersions for any galaxy, we used the fixed values of $R_\mathrm{e} = 30\arcsec$ and $\sigma_g$ = $\unit[50]{km\,s^{-1}}$. For the GCs associated with galaxies, the distances were assumed to be the same as their host galaxy, adopting the distance from \citet{2009ApJ...694..556B} and, for the intracluster GCs, the distances were taken as \unit[16.5]{Mpc} \citep{2007ApJ...655..144M}. 

We report four GCs with nebular emission lines detected through this workflow. The spectra of these candidates are shown in Fig.~\ref{fig:spec}. We elaborate on their properties in the following paragraphs.

\textit{GC-a} (R.A.=$12\fh28\fm39\fs69$, Decl.=$+7\arcdeg53\arcmin32\farcs96$). Also known as RZ 2109, this GCs was discovered to host the X-ray signature of an accreting stellar black hole \citep{2007Natur.445..183M}. Follow-up studies confirmed the existence of strong and broad $\lambda 4959$, $\lambda 5007$ emission \citep{2008ApJ...683L.139Z, 2012ApJ...759..126P} and attributed this emission to the photoionization of a strong wind driven from a stellar-mass black hole.

\textit{GC-b} (R.A.=$12\fh29\fm44\fs51$, Decl.=$+8\arcdeg0\arcmin29\farcs67$). This cluster is classified as old ($\sim\unit[10]{Gyr}$) and metal-poor ($Z\sim0.004$) based on the comparison to the BC03 models. The radial velocity of the cluster is $\unit[970\pm11]{km\,s^{-1}}$, associated with M49. It has a solid detection of an emission feature around \unit[5007]{\AA} with a significance of $\sim\unit[9]{\sigma}$ while H$\beta$ and $\lambda 4959$ are doubtful. Its absolute magnitude $M_{5007}$ is around $\unit[-3.1]{mag}$, within the range of the PNLF. However, the center of the emission line is not aligned with \unit[5007]{\AA} and has an offset of \unit[2.3]{\AA} in the rest frame (corresponding to $\Delta v_\mathrm{r} \sim \unit[140]{km\,s^{-1}}$), which casts doubt on its association with the GC. We adopted the same criterion as \citet{2013ApJ...769...10J} that the difference of radial velocity, $\Delta v_\mathrm{r}$, between emission lines (coming from the star) and absorption lines (coming from the cluster) should satisfy this relation $\Delta v_\mathrm{r} \leqslant 3\sigma_\mathrm{eff}$, where $\sigma_\mathrm{eff}$ is the combination of the system's internal velocity dispersion and the uncertainty in radial velocity measurement. Its internal velocity dispersion was assumed to be a typical value of $\unit[10]{km\,s^{-1}}$ and the uncertainties of the radial velocity of the cluster and the emission line are $\unit[11]{km\,s^{-1}}$ and $\unit[7]{km\,s^{-1}}$, respectively. The overall $\sigma_\mathrm{eff}$ is estimated to be $\sigma_\mathrm{eff} \approx \unit[17]{km\,s^{-1}}$. Therefore, we ruled out its association of this emission line with a GC. Given it's close distance to the center of M49 \citep[43\farcs5, ][]{2006ApJS..164..334F}, we attributed its source to be a superposition with a field PNe in the M49. To estimate the likelihood, we calculated the possibility of a field PNe being detected by one fiber in our survey. We adopted the best fitting core-S{\'e}rsic surface brightness profile \citep{2003AJ....125.2951G} for M49 from \citet{2006ApJS..164..334F}. The number of PNe is connected with the underlying stellar population through $\alpha$:
\begin{equation}
    N_\mathrm{PN} = \alpha L_\mathrm{bol}
\end{equation}
where the $L_\mathrm{bol}$ represents the bolometric luminosity. The bolometric correction is done via
\begin{equation}
    I = 10^{-0.4(\mathrm{BC}_{V}-\mathrm{BC}_\odot)}10^{-0.4(\mu-K)}
\end{equation}
with the solar bolometric correction $\mathrm{BC}_\odot=-0.07$, and $K=\unit[26.4]{mag\,arcsec^{-2}}$. $\mathrm{BC}_{V}$ is assumed to be fixed, $\mathrm{BC}_{V} = -0.85$, with 10\% accuracy \citep{2006MNRAS.368..877B}. As for the $\alpha$, we adopted the value from \citet{2017A&A...603A.104H} who conducted a photometric survey for PNe in the extended halo of M49. Even though their survey excluded objects within a major-axis radius of \unit[159]{\arcsec} and $\alpha$ might be lower close to the center, it is still a reasonable value to estimate the contribution from field PNe. We calculated the expected number of field PN in the aperture of each fiber (1\farcs5 in diameter) by coincidence and derived the possibility of having detected one PN in our survey to be around 20\%, suggesting it is possible to detect a field PN by chance.

\textit{GC-c} (R.A.=$12\fh24\fm30\fs69$, Decl.=$+12\arcdeg54\arcmin3\farcs94$). The peculiar feature of this cluster is that it has double-peaked emission lines at both \unit[5007]{\AA} and \unit[4959]{\AA}. The red peaks lie at the exact wavelength of [O{\sc iii}] while the blue peaks are shifted by $\sim$\unit[2.4]{\AA} ($\unit[140]{km\,s^{-1}}$), which is larger than $3\sigma_\mathrm{eff}$. Given its distance from the center of M84 ($\sim\unit[8]{arcmin}$), it's unlikely to be a field PN superposed along the line of sight. A possible source is H{\sc ii} region associated with the NGC 4438-N4388-M86 complex, similar to what was found by \citet{2002ApJ...580L.121G}. \citet{2005A&A...437L..19O} found evidence of an H{\sc i} cloud therein, suggesting neutral gas stripped from the cluster galaxies. Although the cluster is close to the extended ionized gas region \citep[see Fig.~9 in ][]{2018A&A...614A..56B}, there is no sign of H$\beta$ or H$\alpha$ emission even after we stacked the spectra of surrounding clusters. In that case, it is hard to imagine that these two peaks, which have almost the same flux, come from completely different physical processes. It has been suggested that a double-peaked line profile may result from the expanding shell of a PN \citep[e.g.,][]{1998A&A...329..265G}. \citet{2010A&A...523A..86S} argued that the peak line emission corresponds to the denser inner regions where the emission is high. In their models, the expansion rates derived from the line peak separations, have a weak dependence on the metallicity and may reach $\unit[50]{km\,s^{-1}}$. Despite the fact this value is smaller than what we found in this cluster ($\unit[70]{km\,s^{-1}}$), it is possible that this emission comes from a fast expanding PN. Given that this process is still poorly understood and, for the sake of clarity, we excluded this candidate in the future analysis.

\textit{GC-d}. This cluster is located at right ascension $12\fh32\fm17\fs72$, declination $+12\arcdeg06\arcmin25\farcs42$ (J2000), and has a radial velocity of $\unit[1328\pm34]{km\,s^{-1}}$. The best fit model for this cluster has metallicity of $Z=0.0004$ and age of \unit[10]{Gyr}. [O{\sc iii}] $\lambda 5007$ is well-measured ($7\sigma$), [O{\sc iii}] $\lambda 4959$ is marginally detectable ($2.4 \sigma$) and no emission is found at H$\beta$ and H$\alpha$. The flux ratio between $\lambda 4959$ and $\lambda 5007$ is close to one third. This cluster lies far away enough from any galaxy, thus the likelihood of being a superposed field PN is negligible. Since there is no ionized region close to this cluster, we attribute the emission to be from a PN. The radial velocity determined from the [O{\sc iii}] emission lines is close to that from the absorption lines within the uncertainties, suggesting an association with the GC.

To sum up, of the four candidates with emission around \unit[5007]{\AA}, we found one GC whose emission lines can be attributed to a PN. The flux ratio $I(\lambda 4959)/I(\lambda 5007)$ and radial velocity all support that the emission comes from a PN within this GC. Therefore, we report a single detection out of 1469 GCs in the Virgo cluster.
 
\section{Discussion \label{sec:discussion}}
\subsection{Cut-off magnitude $M_*$ \label{sec:cutoff}}
In this work, we adopted the standard luminosity function of PNe from \citet{1989ApJ...339...53C}\footnote{But see also \citet{1993A&A...275..534M, 2008ApJ...681..325M, 2013A&A...558A..42L, 2015A&A...575A...1R} for other representations of PNLF.}:
\begin{equation}
N(M) \propto e^{0.307M}\left(1-e^{3\left(M_*- M\right)}\right)
\end{equation}
where $M_*$ is the cut-off magnitude of the PNLF. For stellar populations more metal-rich than the Large Magellanic Cloud (LMC), they are confirmed to have the same value of the cut-off $M_* \sim \unit[-4.5]{mag}$ \citep[e.g.,][]{2012Ap&SS.341..151C}. However, as the population becomes more metal-poor, it has been argued that the cut-off depends on the metallicity \citep{1992ApJ...389...27D, 2010A&A...523A..86S}. The detailed formation and evolution of a PN is regulated by the metallicity in some aspects. As the metallicity decreases, the cooling efficiency of the gas drops off, causing a higher electron temperature. It can also reduce the power of the stellar wind from the CS, which leads to a more dilute environment compared with their metal-rich counterparts. The radiation field and the luminosity of the CS are also influenced by the metal content. Based on the photoionization/radiation-hydrodynamics in PNe models, both \citet{1992ApJ...389...27D} and \citet{2010A&A...523A..86S} found the [O{\sc iii}] emissivity has a strong dependence on the metallicity, i.e., the cut-off magnitude becomes fainter as the metallicity decreases, although there is some discrepancy towards the metal-rich end.

This dependence on metallicity is, to some extent, confirmed by the observations of extragalactic systems with known distances. By collecting the distance of 16 galaxies derived from Cepheid and Tip of the Red Giant Branch (TRGB) measurements, \citet{2012Ap&SS.341..151C} found that $M_*$ does fade in metal-poor populations (see their Fig.~5), roughly following the prediction of \citet{1992ApJ...389...27D, 2010A&A...523A..86S}. But we remind the reader of the large measurement uncertainties in these metal-poor and low-mass systems due to the low number of PNe and poorly defined luminosity functions. The other caveat that may limit its validity for GCs is that the sample of \citet{2012Ap&SS.341..151C} only extends to SMC-like metal-poor systems, which is still more metal-rich than typical GCs ([Fe/H]$ = -1$) by \unit[0.5--1]{dex}. If we are willing to extrapolate this relation \citep[Eq.~4.1 in][]{1992ApJ...389...27D} to the lower metallicities, the cut-off is expected to be at least \unit[1]{mag} fainter. However, there is one PN born in a metal-poor environment that may not follow this prediction. As shown in Table~\ref{tab:pngc}, Ps 1 is a confirmed PN found in an extremely metal-poor GC. The expected cut-off for such a population is around $M_*=\unit[0.6]{mag}$, which is marginally fainter than the observed magnitude of [O{\sc iii}] emission from this PN. This finding tentatively casts some doubt on the robustness of these models (at least in GC-like systems) and calls for a better theoretical explanation for the PNLF itself.

As regards to our sample in the Virgo cluster, $M_{5007}$ of the PN in GC-d (and GC-c if included) is \textit{brighter} by \unit[1.5--2]{mag} than the cut-off predicted by \citet{1992ApJ...389...27D}'s model. Nevertheless, as discussed in Sec.~\ref{sec:find}, there is no explanation for the [O{\sc iii}] emission other than a PN in at least one cluster. Therefore, we do not rule out the existence of a PN in GC-d based on this fainter cut-off magnitude, but rather take it as the evidence to question the steepness of this correlation in extremely metal-poor environments. The binary formation channel is likely to dominate in a GC-like environment (see Sec.~\ref{sec:binary}) and these coalesced binary stars, such as blue straggler stars, have been proposed to account for the brightest of PNe \citep{2005ApJ...629..499C}. Unfortunately, none of the currently available models (including \citet{1992ApJ...389...27D, 2010A&A...523A..86S}) include binaries to explain the cut-off of the PNLF. This may explain why the models fail to predict the cut-off in a GC-like population correctly, and our work, although just a single detection, might be helpful in constraining the evolution model of PN within the framework of binary interaction.

To provide a closer look at the influence of the cut-off, we consider two choices for the cut-off in the following analysis: $M_*=\unit[-4.5]{mag}$ and $M_*=\unit[-4.1]{mag}$ to represent the case whether the cut-off has a dependence on metallicity. The latter one is chosen to be the magnitude of the brightest PN in our sample.

\subsection{Detection Limit \label{sec:detection}}

Before we draw any conclusion from this single detection, it is crucial to assess the significance of this finding. Compared with previous extragalactic surveys for PNe \citep{2012ApJ...752...90P, 2013ApJ...769...10J}, this apparent deficiency (1 out of 1469) of PNe could be intrinsic properties of Virgo GCs, or because most PNe are below our observational detection limits. A cluster whose spectrum is not good enough to detect even the brightest PNe should have zero weight in the estimation of the luminosity-specific frequency of PN. Therefore, we use the detection limit, i.e., the minimum emission flux required for detection for an individual GC's spectrum, to quantify the contribution of a single GC to the estimation of $\alpha$.

\begin{figure*}[ht!]
\plotone{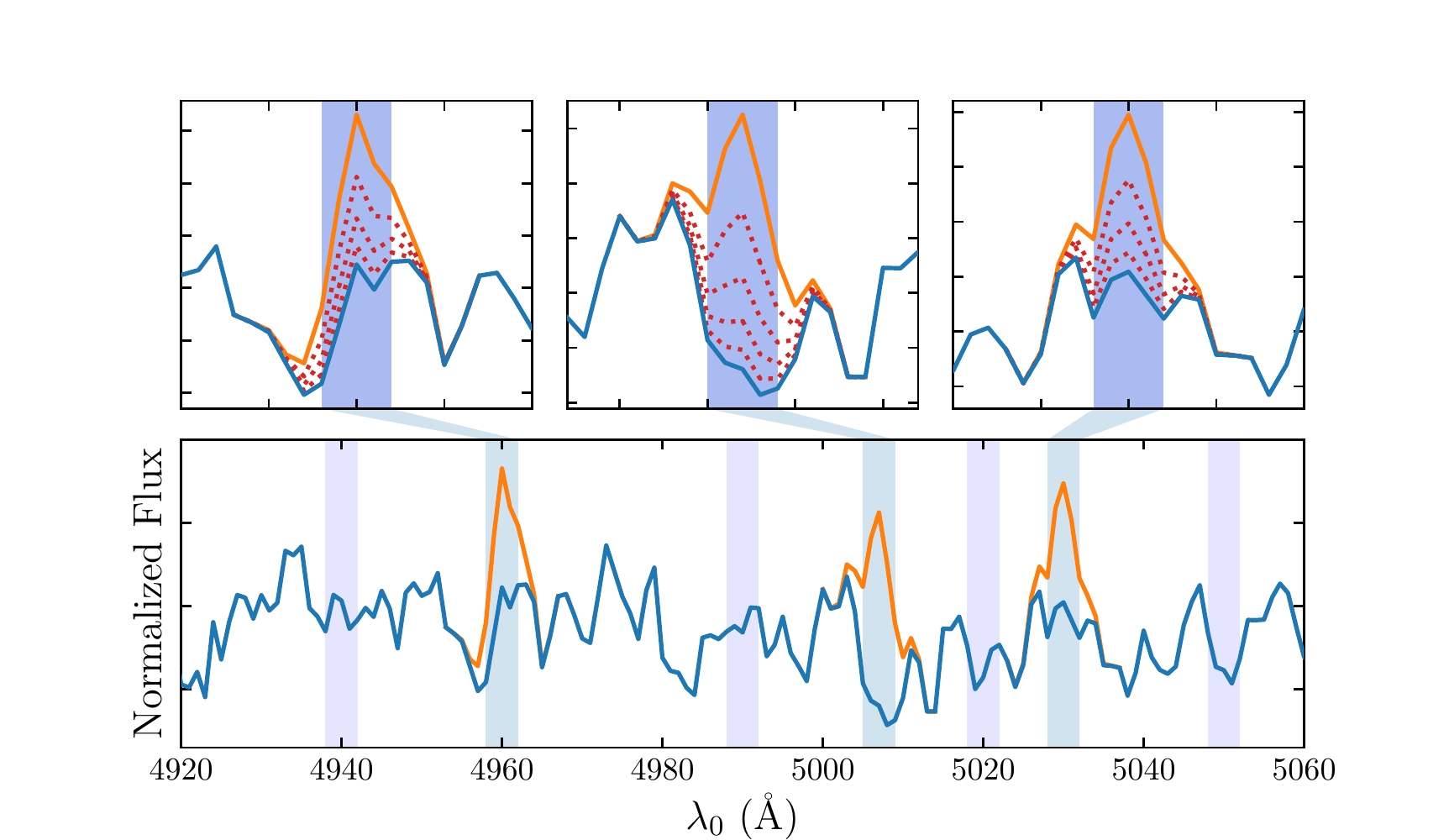}
\caption{Mock line tests. A Gaussian-like profile is added to the calibrated spectrum at the specific wavelength. The intensity of the mock line increases until the new line can be detected by the code. The above process is applied at various wavelengths (shown as the blue boxes) to eliminate random effects. Zoomed-in spectra are shown at the top. Red dotted curves represent the spectra with mock lines added. The mock line profiles are marked orange when they are detected. The average value of the intensity needed for detection is taken as the detection limit for each individual observation. \label{fig:mock_line}}
\end{figure*}

To estimate the detection limit, we added a few mock lines around \unit[5007]{\AA} to check the intensity required to get detected. A Gaussian profile with $\sigma=\unit[1.8]{\text{\AA}}$, corresponding to the spectral resolution, was used to simulate a mock emission line. Fig.~\ref{fig:mock_line} exhibits one mock line test with zoomed-in views at the top. The central wavelength of the mock line was shifted by several angstroms (shown as the blue boxes) to eliminate random effects (possible emission/absorption feature, uncertainty in radial velocity determination or detection failure). In the zoomed-in figures, the intensity of mock lines is increased by a factor of 1.5, corresponding to red dotted curves. The mock lines are marked orange when they are detected. We then took the average value of the intensity needed for detection as the detection limit for each individual observation and transformed it into absolute magnitude $M_\mathrm{lim}$ using the same method as for $M_{5007}$.

\begin{figure*}[ht!]
\plotone{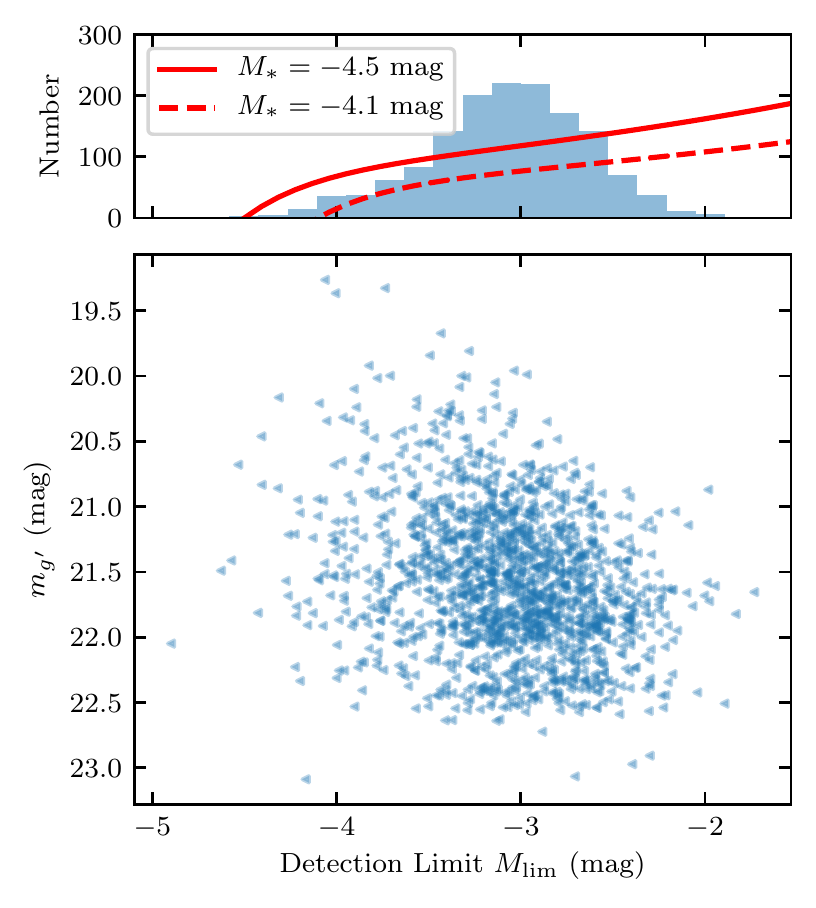}
\caption{Detection limits versus the $g^\prime$ band magnitudes of GCs. There is a rough correlation between the detection limit and the brightness of the cluster, in the sense that brighter clusters have brighter $M_\mathrm{lim}$. In the top panel, the histogram of the detection limits are present, compared with the PNLF of different cut-off: $M_*=\unit[-4.5]{mag}$ (red solid) and $M_*=\unit[-4.1]{mag}$ (red dashed). Any cluster whose detection limit is brighter than the bright-end cut-off cannot have a single PN detected. Only 4 and 27 GCs have detection limits brighter than the cut-off of $M_*=\unit[-4.5]{mag}$ and $M_*=\unit[-4.1]{mag}$ respectively, illustrating the capability of detecting PNe using our data. \label{fig:detection}}
\end{figure*}

Details of the detection limits are present in Fig.~\ref{fig:detection}. Detection limits versus $g^\prime$ band magnitudes of GCs are shown in the bottom panel and we present the histogram of detection limits at the top. We found a rough correlation between the detection limit and the brightness of the cluster, where brighter clusters have brighter $M_\mathrm{lim}$. Most GCs are brighter than the sky in the fiber, therefore we expect that the detection limits for more luminous GCs are brighter due to larger noise. It is noticeable, however, that there is a large scatter, which may result from variation in observing conditions. Only 4 and 27 GCs have detection limits brighter than the cut-off of $M_*=\unit[-4.5]{mag}$ and $M_*=\unit[-4.1]{mag}$ respectively, illustrating the capability of detecting PNe using our data.

\subsection{$\alpha$ for GCs in the Virgo cluster\label{sec:alphaingcs}}

We used Monte Carlo (MC) simulations to constrain the value of $\alpha$ based on the single detection. $\alpha$ represents the specific-luminosity frequency for PNe within the range of the PNLF. The faint end of cut-off of the PNLF is taken to be \unit[8]{mag} below $M_*$ following \citet{1963ApJ...137..747H}'s prediction. $\alpha$ was assumed to be identical for every cluster in our sample. Based on the best fit BC03 model, we derived the bolometric luminosity for each GC and got an expected value for the number of PNe by multiplying the luminosity by $\alpha$. These simulated PNe were assigned with a certain brightness $M_{5007}$ following the PNLF. Then, we counted the number of clusters with simulated PNe which have $M_{5007}$ beyond their detection limits. Since we had one GC detected with a PN, we calculated the probability $p(\theta|\alpha)$ of a single-detection $\theta$ for a given $\alpha$ by repeating the above procedures 1000 times. Using Bayes' Theorem:
\begin{equation}
p(\alpha|\theta) = \frac{p(\theta|\alpha)p(\alpha)}{p(\theta)}
\label{eq:bayes}
\end{equation}
and assuming the prior to be a uniform distribution, we can calculate the probability distribution of $\alpha$ given a single detection ($p(\alpha|\theta)$).

\begin{figure*}[ht!]
\plottwo{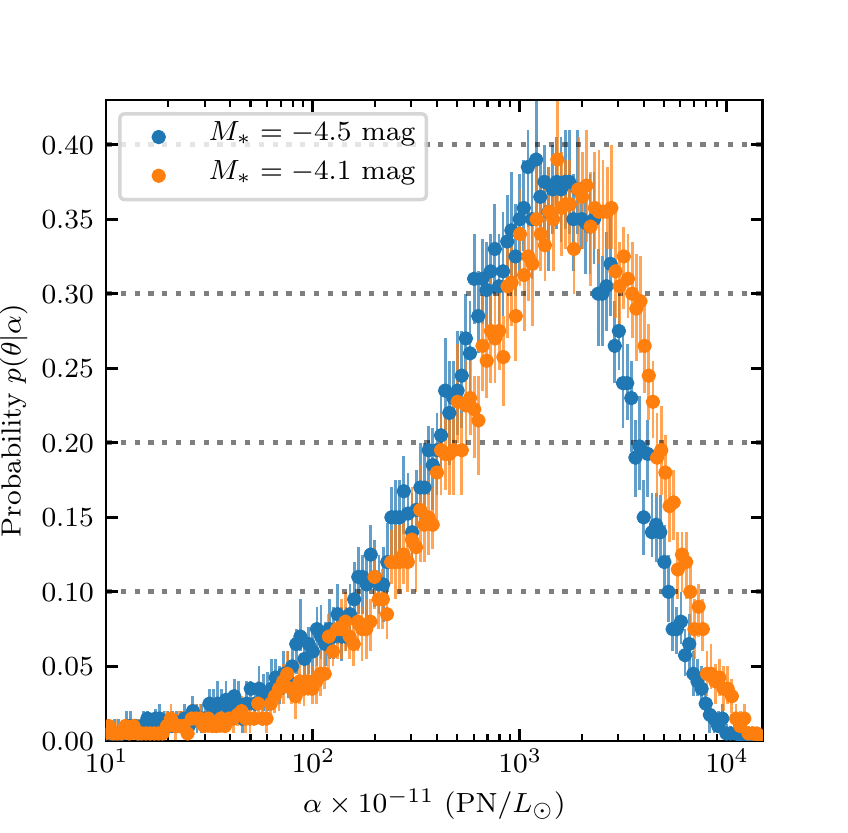}{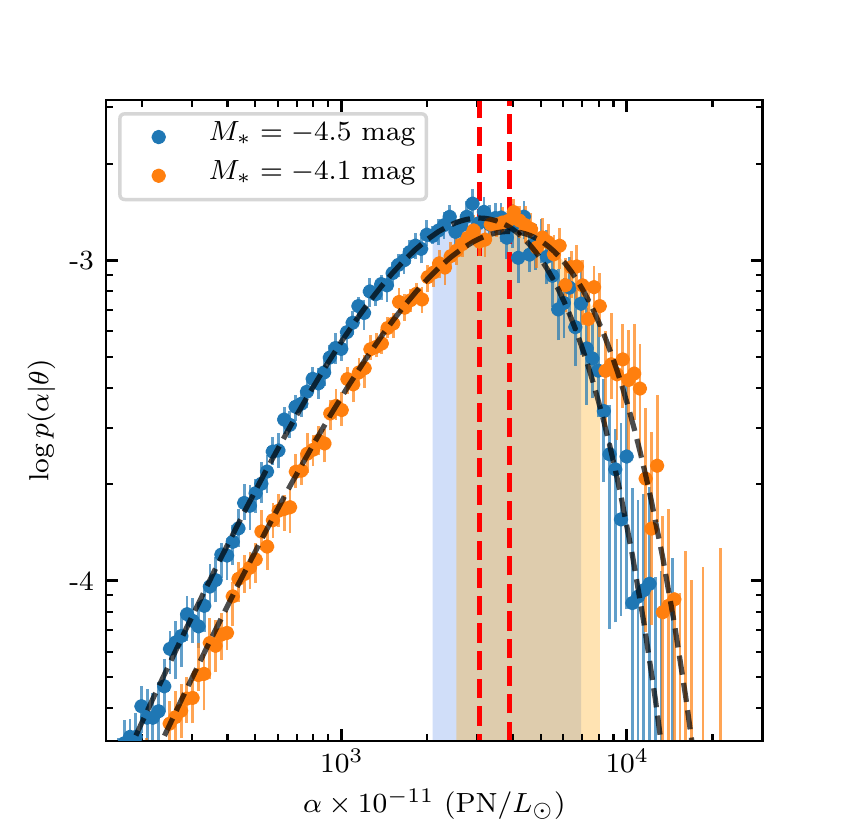}
\caption{Probability of a single detection for various $\alpha$ on a log scale based on Monte Carlo (MC) simulations with different cut-off magnitude: $M_*=\unit[-4.5]{mag}$ (blue) and $M_*=\unit[-4.1]{mag}$ (orange). Each MC test was run 1000 times. Left: the probability distribution of a single detection at a given $\alpha$ ($p(\theta|\alpha)$). Right: the probability distribution of $p(\alpha|\theta)$ calculated through Eq.~\ref{eq:bayes}. The red vertical dashed lines represent the maximum-likelihood estimation ($\unit[3.0\times 10^{-8}]{PN/\mathit{L}_\odot}$ for $M_*=\unit[-4.5]{mag}$ and $\unit[3.9\times 10^{-8}]{PN/\mathit{L}_\odot}$ for $M_*=\unit[-4.1]{mag}$) and the shadowed areas represent the 68.3\% confidence interval (from $\unit[2.0\times 10^{-8}]{PN/\mathit{L}_\odot}$ to $\unit[6.6\times 10^{-8}]{PN/\mathit{L}_\odot}$ for $M_*=\unit[-4.5]{mag}$ and from $\unit[3.2\times 10^{-8}]{PN/\mathit{L}_\odot}$ to $\unit[9.1\times 10^{-8}]{PN/\mathit{L}_\odot}$ for $M_*=\unit[-4.1]{mag}$). The black dashed lines indicate the best fit gamma distribution. \label{fig:mc_alpha}}
\end{figure*}

We show the distribution of $p(\theta|\alpha)$ and $p(\alpha|\theta)$ based on Monte Carlo simulations in Fig.~\ref{fig:mc_alpha}. In the left figure, we can see that as $\alpha$ increases, $p(\theta|\alpha)$ first increases from zero and reaches a maximum around 0.35 at $\alpha \sim \unit[10^{-8}]{PN/\mathit{L}_\odot}$. The probability then decreases since more than one simulated PNe are detected in the GCs. In the right panel, the distribution of $p(\alpha|\theta)$ is presented in log-log space with the best fit gamma distribution (dashed line) and the 68.3\% confidence interval (shadowed area). The maximum-likelihood estimations of $p(\alpha|\theta)$ are around $\unit[3.0\times 10^{-9}]{PN/\mathit{L}_\odot}$ and $\unit[3.9\times 10^{-8}]{PN/\mathit{L}_\odot}$, with the 68.3\% confidence interval from $\unit[2.0\times 10^{-9}]{PN/\mathit{L}_\odot}$ to $\unit[6.6\times 10^{-9}]{PN/\mathit{L}_\odot}$ from $\unit[3.2\times 10^{-8}]{PN/\mathit{L}_\odot}$ to $\unit[9.1\times 10^{-8}]{PN/\mathit{L}_\odot}$ for $M_*=\unit[-4.5]{mag}$ and $M_*=\unit[-4.1]{mag}$, respectively. If we include the cluster with double-peaked emission lines, GC-c, $\alpha$ is estimated to increase by a factor of 1.5. See Table~\ref{tab:alpha} for a summary of $\alpha$ in this work.

\begin{deluxetable*}{l|CC}
\tablecaption{$\alpha$ for Virgo GCs and UCDs \label{tab:alpha}}
\tablehead{
    \colhead{$\alpha$ ($\unit{PN/\mathit{L}_\odot}$)} & \colhead{$M_*=\unit[-4.5]{mag}$} & \colhead{$M_*=\unit[-4.1]{mag}$}
}
\startdata
GC (single detection) & 3.0_{-1.0}^{+3.6}\times 10^{-8} & 3.9_{-0.7}^{+5.2}\times 10^{-8} \\ 
GC (two detections) & 4.6_{-1.4}^{+4,5}\times 10^{-8} & 5.9_{-1.8}^{+6.4}\times 10^{-8} \\                                              
UCD (non-detection)   & \leqslant 2.2\times10^{-7} & \leqslant 3.0\times10^{-7} \\
\enddata
\end{deluxetable*}

\citet{2012ApJ...752...90P} reported a non-detection result in M49 and they estimated the upper limit of $\alpha$ by combining the overall bolometric luminosity of all the GCs in their sample which gave a value of $\alpha\leqslant\unit[8\times10^{-8}]{PN/\mathit{L}_\odot}$ \citep[the upper limit is based on the 90\% confidence level for a non-detection;][]{1991ApJ...374..344K}. This upper limit is consistent with our result. Nevertheless, we point out their approach will underestimate the upper limit of $\alpha$ since the GCs with poor quality (bright detection limits) are also included in the total luminosity.

\subsection{$\alpha$ for UCDs in the Virgo cluster}
Following the same procedures as for GCs, we also estimate $\alpha$ for UCDs in the Virgo cluster. UCDs are compact stellar systems that are brighter than most GCs but fainter than compact ellipticals. \citet{2012MNRAS.425..325F} revealed that the UCDs are old and (generally) metal rich (mean [Fe/H] =$-0.8 \pm 0.1$). The catalog of UCDs, which consists of 121 UCDs, was compiled from \citet{2015ApJ...812...34L, 2015ApJ...802...30Z} and Liu et al. (in prep.). After subtracting the best fit SSP models from the spectra, we found no trace of [O{\sc iii}] emission lines in the UCDs. Then, based on the non-detection result, we calculated the probability distribution of $\alpha$ for UCDs using the same method described in Sec. \ref{sec:alphaingcs} and estimated its upper limit to be $\alpha\leqslant\unit[2.2\times10^{-7}]{PN/\mathit{L}_\odot}$ and $\alpha\leqslant\unit[3.0\times10^{-7}]{PN/\mathit{L}_\odot}$ at 68.3\% significance level for $M_* = \unit[-4.5]{mag}$ and $M_* = \unit[-4.1]{mag}$ respectively.

\subsection{Expected value for $\alpha$}

The $\alpha$ for GCs in the Virgo cluster are much lower than any result in the literature, which raises the question of what is expected. Here, $\alpha$ denotes the luminosity-specific PN density within the whole range of the PNLF, which follows the relationship \citep{2005ApJ...629..499C}:
\begin{equation}
    \alpha = B \cdot t \cdot f
\end{equation}
where $B$ is the population's luminosity-specific stellar evolutionary flux, $t$ is the mean timescale for the [O{\sc iii}]-bright stage, and $f$ is the percentage of main-sequence stars that evolve into PNe. $B$ is around $\unit[1-2 \times 10^{-11}]{{L_\odot}^{-1}{yr}^{-1}}$ for old stellar populations and is independent of age and metallicity \citep{2006MNRAS.368..877B}. Based on the calculation of energy available to post-AGB stars, they also estimated that the upper limit of the timescale of PNe should be shorter than $\sim \unit[3 \times 10^4]{yr}$ which is its dynamical timescale. Thus, the measurement of $\alpha$ can provide a constraint on the percentage of stars that turn into PNe.
    
If all stars contribute to the PNLF ($f=1$) as the traditional scenario suggests, $\alpha$ should be smaller than $2 \times 10^{-11} \cdot 3 \times 10 ^4 \sim \unit[6 \times 10 ^{-7}]{PN/\mathit{L}_\odot}$. PNe, in reality, may have shorter lifetimes or lower transformation fractions, thus $\alpha$ should be smaller than this maximum. \citet{2006MNRAS.368..877B} demonstrated that the timescale of $t$ should be extended with increasing age, due to the slow evolution of a low mass star. However, it will reach a ceiling if the CSs take too long to get hot enough and ionize the nebular material. They concluded that any star with a CS mass smaller than \unit[0.52]{$M_\odot$} will not be able to form a PN, which is an important limit for old populations like GCs.
    
The estimation mentioned above is based on the assumption of isolated stellar evolution. But it has been proposed that most PNe are generated through some kind of binary interaction \citep{2006ApJ...645L..57S, 2006ApJ...650..916M}. The binary hypothesis simply states that a majority of PNe have formed through a binary-interaction channel. \citet{1995MNRAS.277.1443H} predicted that $38 \pm 4 \%$ of all PNe have been affected by binary interactions, and those from the post-common-envelope evolutionary channel account for one-third. \citet{2006ApJ...650..916M} argued that $\sim 75 \%$ of the Galactic PNe can be ascribed to a common-envelope interaction. More than 40 PNe have been associated with close binaries now \citep{2011apn5.confP.109M}. However, an overly tight binary system may also disrupt the formation of PNe. If the interaction occurs at early phases, such as during the red giant branch phase, the CS cannot ionize the outer layer that has been blown away. So not all binary systems can produce PNe. \citet{2006ApJ...650..916M} argued that the separation limit for systems to form PNe should be between \unit[100]{$R_\odot$} and \unit[500]{$R_\odot$}. Thus, in a GC-like system, $\alpha$ should be close to zero if we consider isolated stellar evolution. The binary interaction scenario may indeed increase the value, but not significantly.

This low but non-zero value of $\alpha$ in Virgo GCs indicates that binary interaction might have an important role in forming PNe in GCs. Of the four Galactic globular clusters (GGCs) hosting PNe, two of them have a very high CS mass \citep{1993ApJ...411L.103H, 2000AJ....120.2044A}, which correspond to main sequence masses of up to three times the cluster turnoff mass \citep{2000A&A...363..647W}. This could come from a mass-transfer binary or a merger formed in a dense GC environment that later becomes a PN. On the other hand, the low-mass CSs of an old population are intrinsically unable to generate any PNe. However, \citet{2017ApJ...836...93J} completed a full HST survey of all PNe in Galactic GCs and they found only the CS of Ps 1 is higher than the predicted white dwarf mass, thus requiring a history of mass augmentation (the CS of JaFu 1 shows compact and bright [O{\sc iii}] and H$\alpha$ emission, suggesting a binary companion).

\begin{deluxetable*}{lccccr}
\tablecaption{Planetary Nebulae in Globular Clusters \label{tab:pngc}}
\tablewidth{0pt}
\tablehead{
    \colhead{Name} & \colhead{Age\tablenotemark{a} (GC)} & \colhead{$M_V$\tablenotemark{b} (GC)} & \colhead{[Fe/H]\tablenotemark{b} (GC)} & \colhead{$M_{5007}$ (PN)} & \colhead{Reference} \\
    \colhead{Host Galaxy/GC/PN} & \colhead{Gyr} & \colhead{mag} & \colhead{dex} & \colhead{mag} & \colhead{}
}
\startdata
    MW/M15/Ps 1 (K 648) & Old & $-9.19$ & $-2.37$ & $0.49$ & \citet{1928PASP...40..342P}\\
    MW/M22/MRAS 18333 & Old & $-8.50$ & $-1.70$ & $5.04$ & \citet{1989ApJ...338..862G} \\
    MW/Pal 6/JaFu 1 & Old & $-6.79$ & $-0.91$ & $-2.8$ & \citet{1997AJ....114.2611J} \\
    MW/NGC 6441/JaFu 2 & Old & $-9.63$ & $-0.46$ & $2.74$ & \citet{1997AJ....114.2611J} \\
        NGC 3379/gc771/PN & ... & $-8.65$ & $-1.40$ & $-1.62$ & \citet{2006AA...448..155B} \\
        Fornax-dSph/H5/PN & ... & $-7.40$ & $-1.73$ & $1.33$ & \citet{2008AA...477L..17L} \\
    NGC 7457/GC7/PN & ... & $-8.20$ & $-0.40$ & $-3.74$ & \citet{2008AJ....136..234C} \\
    M 31/B115-G117/PN & 14 & $-8.54$ & $0.1 \pm 0.1$ & $-2.0$ & \citet{2013ApJ...769...10J} \\
    M 31/BH16/PN & Old & $-6.46$ & $-1.0$ & $0.7$ & \citet{2013ApJ...769...10J} \\
    M 31/NB89/PN\tablenotemark{c} & $10.4$ & $-6.60$ & $-0.6$ & $0.5$ & \citet{2013ApJ...769...10J} \\
    M 87/GC-d/PN & 10 & $-9.51$\tablenotemark{d} & $-1.71$ & $-4.1$ & this work\\
\enddata
\tablenotetext{a}{Ages were estimated by \citet{2009AJ....137...94C, 2011AJ....141...61C}.}
\tablenotetext{b}{For the Galactic GCs, these data were obtained from the Harris catalog 2010 edition \citep{1930HarMo...2.....S, 1979ARAA..17..241H, 1996AJ....112.1487H}.}
\tablenotetext{c}{\citet{2015AJ....149..132B} ruled out this one hosting a PN.}
\tablenotetext{d}{The transformation relation is adopted from \citet{2013AA...552A.124B, 2005AJ....130..873J}.}
\end{deluxetable*}

In Table~\ref{tab:pngc}, we listed all of the GC PNe that are currently published. The PNe span a wide range of [O{\sc iii}] luminosities, from close to the cut-off of the PNLF to \unit[9]{mag} fainter. The PN we found in M87 is the brightest one ever found in a GC. \citet{2013ApJ...769...10J} noted that two GGCs with a PN also have a Low-mass X-ray binary (LMXB) embedded. Since the presence of an LMXB is connected to tight binary systems, this may suggest a possible binary origin for the GC PNe. However, by analyzing the stellar interaction rate in the dense cores of these clusters, \citet{2012ApJ...752...90P} concluded no strong evidence exists. We matched our sample with the catalogs of known GCs that host LMXBs in the Virgo cluster \citep{2003ApJ...586..814M, 2004ApJ...613..279J, 2008ApJ...689..983H, 2018ApJ...862...73L} and found 17 GCs containing a LMXB. Given the field of view of \textit{Chandra} X-ray observations, previous studies only targeted at the very center of the galaxies, which covers a fraction ($\le 5\%$) of our sample. None of the GCs discussed in Sec.~\ref{sec:find} were observed in these studies.

\subsection{$\alpha$ in Globular Clusters and Galaxies \label{sec:binary}}
\begin{figure*}[ht!]
\plotone{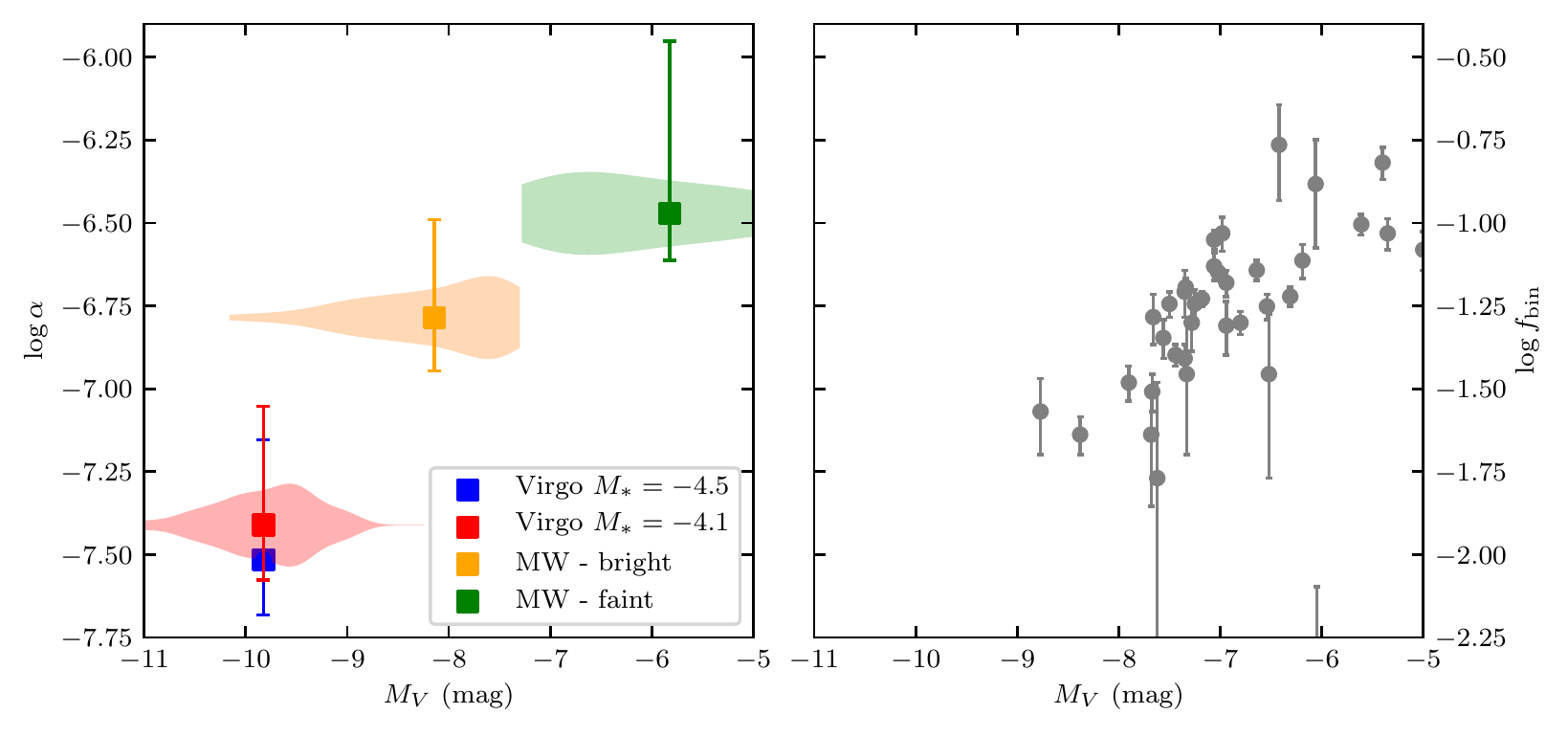}
\caption{Left: Number of PNe per bolometric luminosity $\alpha$ versus absolute $V$ band magnitude of Virgo GCs ($M_* = \unit[-4.5]{mag}$, blue, and $M_* = \unit[-4.1]{mag}$, red)), GGCs brighter than the turnover luminosity ($M_V=\unit[-7.3]{mag}$, orange) and GCCs fainter than the turnover luminosity (green). The values of $\alpha$ in GGCs do not rely on the choice of different cut-offs. The thickness of the horizontal violin plots represent the distribution of the brightness of GCs in each group. The error bars represent the 68.3\% confidence intervals of $\alpha$ for each group. Because the two different cut-off magnitudes yield similar results for $\alpha$, in the following analysis, we use the more conservative choice of $M_*=\unit[-4.1]{mag}$. Right: Fraction of binaries with $q \ge 0.5$ in the core \citep{2012A&A...540A..16M} as a function of the absolute visual magnitude of the host GC in the Milky Way. Note that the scales of $\alpha$ and $f_\mathrm{bin}$ are exactly the same. $\alpha$ decreases as GCs becoming brighter, suggesting $\alpha$ is influenced by an environmental factor in cluster and the binary fraction could be the key to understand this correlation. \label{fig:mv_alpha}}
\end{figure*}

The $\alpha$ in GGCs is calculated to be $\alpha=\unit[2.0_{-0.6}^{+1.5}\times10^{-7}]{PN/\mathit{L}_\odot}$ by summing the bolometric luminosity of GCs. The combined luminosity was estimated through the absolute $V$ band magnitude from \citet{1996AJ....112.1487H} with bolometric corrections. Considering \citet{1997AJ....114.2611J}'s survey went \unit[9]{mag} fainter than the cut-off, it is reasonable to assume the sample in \citet{1997AJ....114.2611J} to be complete even for the faintest PNe. Note that $\alpha$ for GGCs calculated in this method has no dependence on the choice of cut-off magnitudes. By comparing with that in GGCs, we found $\alpha$ in the Virgo cluster GCs is around 5--6 times smaller.

The binary fraction in the GCs could be the key to understand the difference. Since most GCs in our sample have $g^\prime$ band magnitude of $ m_{g^\prime} < \unit[22.5]{mag}$, which is \unit[2--3]{mag} brighter than the expected turnover of the GC luminosity function \citep[GCLF,][]{2001stcl.conf..223H} at the distance of Virgo, they are more massive and denser compared with their Galactic counterparts. The ``fundamental plane'' of GCs shows that their size (half-light radius) has no dependence on their mass \citep{2000ApJ...539..618M}, which is also confirmed in Virgo cluster GCs \citep{2005ApJ...634.1002J}. An anti-correlation between cluster binary fraction with its absolute magnitude has been found in GGCs \citep{2012A&A...540A..16M}. Therefore the Virgo GCs in our sample might be expected to have lower binary fractions, which may suggest a reduced efficiency in forming PNe if they are formed through a binary interaction channel. To illustrate this effect, we divided GGCs into two groups: those brighter than the turnover luminosity and those fainter. Based on the detection of PNe in GGCs (three in bright GGCs and one in faint GGCs, see Table \ref{tab:pngc}), we calculated $\alpha$ for the two groups, respectively. The results, together with Virgo GCs are shown in the left panel of Fig.~\ref{fig:mv_alpha} with different colors denoting different groups. Because the two different cut-off magnitudes yield similar results for $\alpha$, in the following analysis, we use the more conservative choice of $M_*=\unit[-4.1]{mag}$.

We see a trend that $\alpha$ decreases for brighter GCs, showing a similar tendency as the binary fraction (right panel). The effect of distance difference between Virgo GCs and GGCs has been taken into consideration in the Sec.~\ref{sec:detection} through the detection limit analysis. This suggests a connection between the efficiency of forming PNe in GCs with the binary fraction. Massive GCs with high stellar density may suppress the binary fraction due to the effect of binary disruption and therefore inhibit the formation of PNe from the binary interaction channel. Another possible cause might be ram-pressure stripping by the intracluster medium. \citet{2003ApJ...585L..49V} shown that ram pressure can effectively alter the morphology of PNe and most of the mass ejected during the AGB phase is left downstream. However, we argue that this is probably not the underlying cause for two reasons. First, there are plenty of PNe outside of GCs in these massive galaxies, and the $\alpha$ measured in the field is higher than in GCs (see the following text and Fig.~\ref{fig:metal_alpha}). This is the reverse of what one might expect, as the field stars are more centrally concentrated than the GCs. Second, M49's $\alpha$ in the field is lower than M87's, which is the opposite of naive expectations for a ram-pressure mechanism. Another possible cause is an internal mechanism within GCs (e.g., collision or nearby stellar radiation) that may shorten the lifetime of PNe by disrupting the stellar ejecta.

\begin{figure*}[ht!]
\plotone{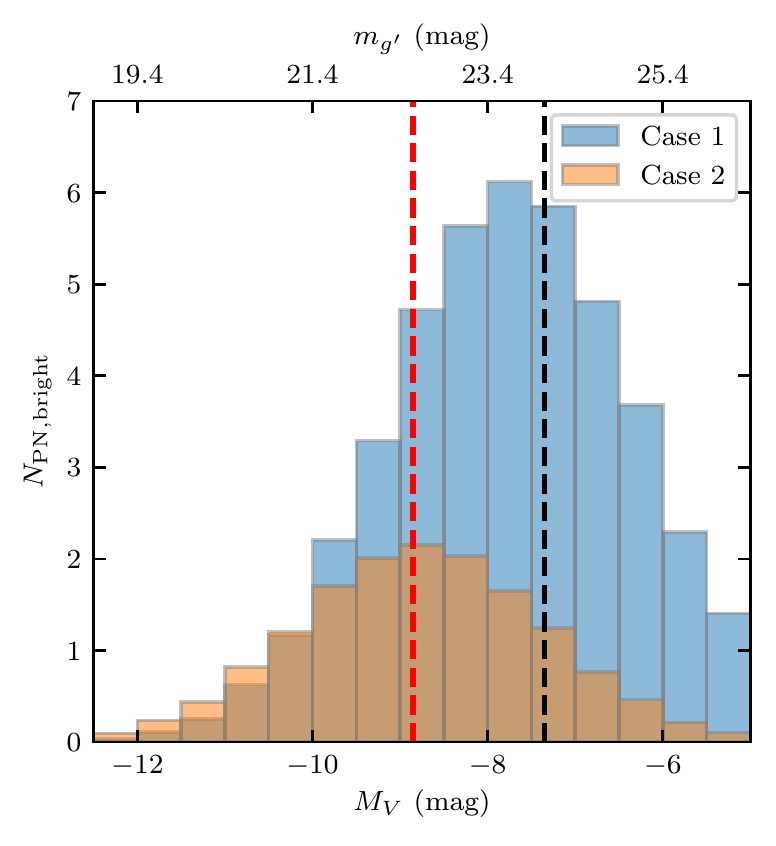}
\caption{Expected number of GCs hosting bright PNe ($M_{5007} < \unit[-2]{mag}$) detected per magnitude bin ($\Delta M_V = \unit[0.5]{mag}$). The blue and orange histograms represent the distribution of the number for case 1 and case 2, respectively. Case 1 adopts a correlation between $\alpha$ and $M_V$ in which $\alpha$ is largely driven by internal processes (binary fraction), while case 2 adopts the same $\alpha$ for all GCs, having no dependence on $M_V$. The red vertical dashed line indicates the faintest GCs targeted for spectroscopy in our survey. There is a significant difference between the expected number of PNe for case 1 and case 2 in a deeper survey. For case 1, if a survey goes deeper by \unit[1.5]{mag} down the GCLF (black vertical dashed line), $\sim 20$ bright PNe will be discovered. For case 2, however, we expect the number to be around 5. A deeper survey could be an effective method to differentiate between these two cases. \label{fig:est}}
\end{figure*}

If the binary fraction is the driving force behind the difference of $\alpha$ in different groups, as suggested by \citet{2005ApJ...629..499C} and \citet{2013ApJ...769...10J}, the formation efficiency of PNe is expected to correlate with the cluster mass or absolute luminosity, $M_V$. This correlation can be represented by a power-law relation between $\alpha$ and $M_V$ with an index of $0.25\pm0.08$. This relationship, however, is only valid if Virgo GCs and GGCs are intrinsically similar, with $\alpha$ largely driven by internal processes. It could instead be the case that the low $\alpha$ in Virgo GCs is due to their unique environment (i.e., in giant ellipticals or in a galaxy cluster), in which case we might expect $\alpha$ in fainter Virgo GCs to be similarly low. Below, we explore how a survey for PNe in fainter GCs can distinguish between these two cases.

Case 1 adopts a correlation between $\alpha$ and $M_V$ as mentioned before, while case 2 considers a fixed $\alpha$ for all GCs, having no dependence on $M_V$. In a flux-limited survey, the effect of the GCLF should also be taken into consideration. Here we adopt a symmetric Gaussian GCLF of $\sigma_{\mathrm GCLF}=\unit[1.37]{mag}$ \citep{2009ApJ...703...42P} and took the total number of GCs around the giant elliptical galaxies M87, M49, M86 and M84 ($29434\pm 1715$) from \citet{2008ApJ...681..197P}. Moreover, since the sensitivity to PNe is mainly determined by the flux level of background (see Fig.~\ref{fig:detection}), sky background becomes the dominant component for noise and the detection limit should remain unchanged as the survey goes deeper. In Fig.~\ref{fig:est}, we present the expected number of GCs hosting PNe detected per magnitude bin ($\Delta M_{V}=\unit[0.5]{mag}$). The blue and orange histograms represent the expected number of candidates per magnitude bin ($\Delta M_V=\unit[0.5]{mag}$) for case 1 and case 2, respectively. The red vertical dashed line indicates the brightness of the faintest GCs targeted for spectroscopy in our survey. 

We find a significant difference between the expected number for case 1 and case 2. For case 1 GCs hosting PNe are most likely to be discovered in clusters with absolute magnitude of $M_{V}\sim\unit[-7.7]{mag}$ ($m_{g^\prime}\sim\unit[23.6]{mag}$) while the value for case 2 is \unit[1.4]{mag} brighter. We predict that for every \unit[0.5]{mag} down the GCLF that is surveyed, there will be more than 2 times more candidates detected for case 1 compared with case 2. For case 1, if a survey goes deeper by \unit[1.5]{mag} down the GCLF, $\sim 20$ PNe brighter than $M_{5007}<\unit[-2]{mag}$. For case 2, however, we expected the number to be around 5. This suggests that this method can differentiate between these two cases effectively.

PNe populations are known to be correlated with their galaxy's properties. More massive, metal-rich galaxies tend to have lower $\alpha$ \citep{1990RPPh...53.1559P, 1993ApJ...414..463H, 2005ApJ...629..499C}. Another anti-correlation is observed with the far-ultraviolet (FUV) excess of a galaxy, where many extreme horizontal branch stars will skip the AGB phase of stellar evolution entirely, reducing the number of PNe generated. \citet{2006MNRAS.368..877B} also pointed out a trend of $\alpha$ with galaxy color --- a poorer PN population in redder ellipticals. 

\begin{figure*}[ht!]
\plotone{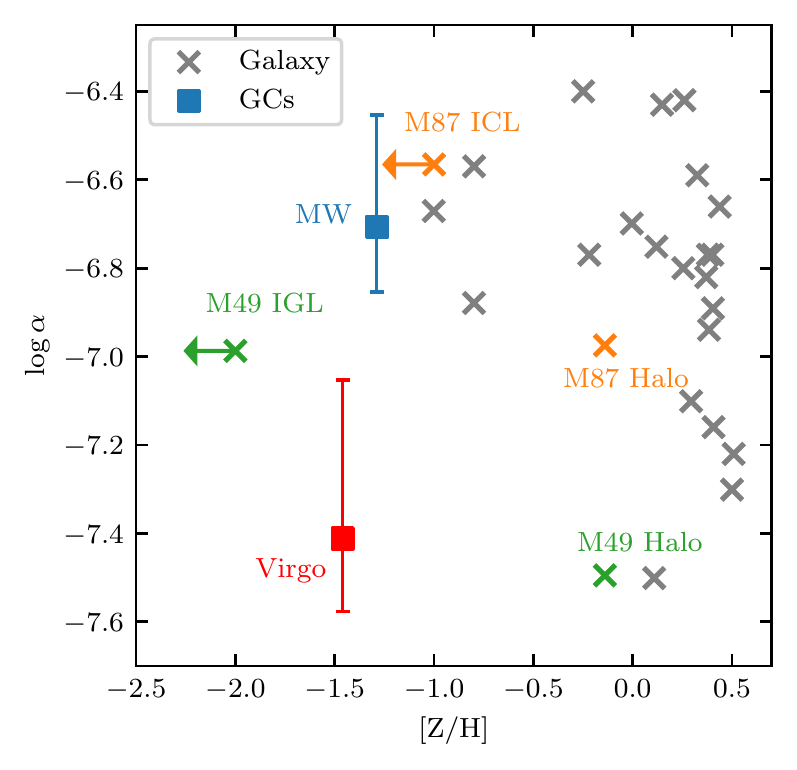}
\caption{Number of PNe per unit bolometric luminosity $\alpha$ as a function of the host stellar population's metallicity. Crosses represent early-type galaxies, the MW bulge, LMC, and SMC. Filled squares show the GC systems. For Virgo GCs, the cut-off is adopted as $\unit[-4.1]{mag}$. Symbols with the same color are from the same galaxy but different environments (GCs, Halo, IGL, ICL). \label{fig:metal_alpha}}
\end{figure*}

Figure \ref{fig:metal_alpha} compares $\alpha$ for the MW's GC system, Virgo GC system, and a sample of galaxies (taken from Table 4 and Table 6 of \citet{2006MNRAS.368..877B} which references $\alpha$ from \citet{1976ApJ...205...74A, 1980ApJS...42....1J, 1993ApJ...414..463H, 1989ApJ...339...53C, 2002AJ....123..269J, 2003MNRAS.346L..62M, 2006MNRAS.368..877B}). The Virgo cluster has already been targeted by a number of PN surveys \citep{1998ApJ...492...62C, 1998ApJ...503..109F, 2004ApJ...615..196F, 2002AJ....123..760A, 2003AJ....125..514A, 2005AJ....129.2585A, 2009A&A...507..621C} to trace the faint diffuse stellar component that are not bound to individual galaxies. The $\alpha$ in the main galaxy halo and intracluster light (ICL) or intragroup light (IGL) of M87 and M49 were from \citet{2015A&A...579A.135L} and \citet{2017A&A...603A.104H}. The metallicities of M87's halo and ICL were adopted from \citet{2015A&A...579A.135L} and those of M49 were from \citet{2017A&A...603A.104H} (upper limit of metallicities in ICL/IGL). The metallicities of GCs in this work were calculated through the GC color-metallicity empirical relation adapted to Virgo galaxies M49 and M87 derived by \citet{2006ApJ...639...95P}. The average values of metallicity were plotted in each GC system. $\alpha$ in four Galactic GCs is consistent with that in early-type galaxies, accompanied by a large uncertainty (due to the low number of known PNe), while in Virgo GC systems $\alpha$ has much lower estimates.

Thanks to a large number of GCs in our sample, we are able to derive a much tighter constraint on $\alpha$ than in previous studies. The correlation between $\alpha$ and the metallicity of the galaxies is clear, however, a turnover may exist towards lower metallicity. Unfortunately, the physics behind this correlation is not very clear. We believe the primary cause is increasing mass loss at high metallicities. \citet{2009A&A...508.1343W} computed stellar evolution models of AGB stars with different metallicities and found the lifetime varies with metallicities, with the longest lifetime occurred for sub-solar metallicity. \citet{2014ApJ...790...22R} also discovered a similar result, that lifetimes of low-mass AGB stars are expected to become shorter at decreasing metallicity ($Z \leqslant 0.008$). These results may suggest that stellar populations with [Fe/H] between $-0.5$ and $-0.2$ may have a larger value of $\alpha$ than their counterparts with higher or lower metallicity with the same luminosity.

\section{Conclusions \label{sec:conclusion}}

We report the largest systematic survey for PNe in GCs. Our survey in the Virgo cluster, mainly around the giant elliptical galaxies M87, M49 and M86, consists of 1469 GCs in total. By subtracting the best fit model from the spectra, we search for emission lines around \unit[5007]{\AA} in the residual spectra. GC-d has an emission line at the exact wavelength of \unit[5007]{\AA} with absolute magnitude $M_{5007}=\unit[-4.1]{mag}$. The flux ratio $I(\lambda 4959)/I(\lambda 5007)$ is close to one third. We confirm that this is a GC hosting a PN, and suggest it could be formed through binary interactions.

The [O{\sc iii}] emission is brighter than the cut-off predicted by \citet{1992ApJ...389...27D} for extremely metal-poor population, suggesting that the dependence on the metallicity could be weaker than was expected. To estimate the specific-luminosity frequency of PNe $\alpha$, we use the detection limit to quantify the contribution of each GC in the calculation of $\alpha$ and Monte Carlo simulations to constrain the probability distribution of $\alpha$ based on a single detection. The probability distribution $p(\alpha|\theta)$ is well fit by a gamma distribution and $\alpha \sim \unit[3.0_{-1.0}^{+3.6}\times 10^{-8}]{PN/\mathit{L}_\odot}$ and $\alpha \sim \unit[3.9_{-0.7}^{+5.2}\times 10^{-8}]{PN/\mathit{L}_\odot}$ for $M_*=\unit[-4.5]{mag}$ and $M_*=\unit[-4.1]{mag}$ respectively, which is at least 5--6 times lower than $\alpha$ in the Galactic GC system. $\alpha$ decreases towards brighter and more massive clusters, sharing a similar trend as the binary fraction in Milky Way GCs \citep[e.g.,][]{2012A&A...540A..16M}. The discrepancy between the Virgo and Galactic GCs can be explained by the observational bias in extragalactic surveys toward brighter GCs. This low but non-zero efficiency is consistent with a binary origin for these PNe in GCs. The importance of CS binarity in the GC environment here is that binary interactions or mergers could form a blue straggler star that later becomes a PN with a more massive CS compared to the turnoff stars in the cluster. We predict the expected number of PNe within GCs to be detected in a deeper survey under two cases (mainly driven by internal processes or by external environment) and show that a deeper survey could be an effective method to differentiate between these two cases. Combined with $\alpha$ reported in other environments (GCs and elliptical galaxies), there may be a turnover of $\alpha$ at sub-solar metallicities.

\acknowledgments We are grateful to the referees for their constructive input. EWP acknowledges support from the National Natural Science Foundation of China under Grant No. 11573002. AL acknowledges the support by the Centre national d'{\'e}tudes spatiales. HXZ acknowledges support from the CAS Pioneer Hundred Talents Program and the NSFC grant 11421303. MGL and YK acknowledge support from the National Research Foundation of Korea (NRF) grant funded by the Korean Government (NRF-2017R1A2B4004632) and the K-GMT Science Program (PID: 14A-MMT003/2014A-UAO-G18) funded through the Korean GMT Project operated by the Korea Astronomy and Space Science Institute (KASI). IC is supported by Smithsonian Astrophysical Observatory Telescope Data Center. IC acknowledges additional support from the Russian Science Foundation grant 17-72-20119 and from the Program of development at M.V. Lomonosov Moscow State University through Leading Scientific School ``Physics of stars, relativistic objects and galaxies''. We thank the staff of the MMT Observatory for their professional support of our program. This research uses data obtained through the Telescope Access Program (TAP), which has been funded by the National Astronomical Observatories, Chinese Academy of Sciences, and the Special Fund for Astronomy from the Ministry of Finance. Based on observations obtained with MegaPrime/MegaCam, a joint project of CFHT and CEA/DAPNIA, at the Canada-France-Hawaii Telescope (CFHT) which is operated by the National Research Council (NRC) of Canada, the Institut National des Sciences de l'Univers of the Centre National de la Recherche Scientifique of France, and the University of Hawaii. Observations reported here were obtained at the MMT Observatory, a joint facility of the University of Arizona and the Smithsonian Institution.

\vspace{5mm}

\software{Astropy \citep{2013A&A...558A..33A}, Matplotlib \citep{2007CSE.....9...90H}, RVSAO \citep{1998PASP..110..934K}}

\end{document}